# On the Observations of the Sun in Polynesia

Sergei Rjabchikov[1]

[1]The Sergei Rjabchikov Foundation - Research Centre for Studies of Ancient Civilisations and Cultures, Krasnodar, Russia, e-mail: srjabchikov@hotmail.com

**Abstract**[2]

The role of the Polynesian sun god Tagaloa has been studied. The Polynesian characters Maui-tikitiki, Tane and Tiki were related to the sun as well. The solar data of Easter Island are essential indeed. The *rongorongo* text on the Santiago staff about the solar eclipse of December 20, 1805 A.D. has been decoded. The Mataveri calendar was probably incised on a rock in 1775 A.D. So, a central event during the bird-man festival was the day of vernal equinox. The priests-astronomers watched not only the sun and the moon, but also some stars of the zodiacal constellations and other bright stars.

## Introduction

On the strength of PMP *gilak* 'shine, glitter,' *gilap* 'lustre, shine,' *dilap* 'to shine, flash,' *kilap* 'lustre, shine,' and *silap* 'sparkle' (ACD) one can reconstruct the root *la* 'shine etc.' I suppose that PPN *la'aa* 'the sun' came from that word. Other reflexes are as follows: PPN *gigila* 'shine, glitter' and *tugi* 'to ignite, light' (POLLEX). Cf. also Maori *ngiha* (< *ngia*) 'fire; to burn,' Hawaiian *niania* 'shining' and Rapanui *ngiingii* 'hot; burning hot' as well as Samoan *'ila* 'to shine, sparkle' [thus, PPN *kila* means 'ditto;' it is my own reconstruction].

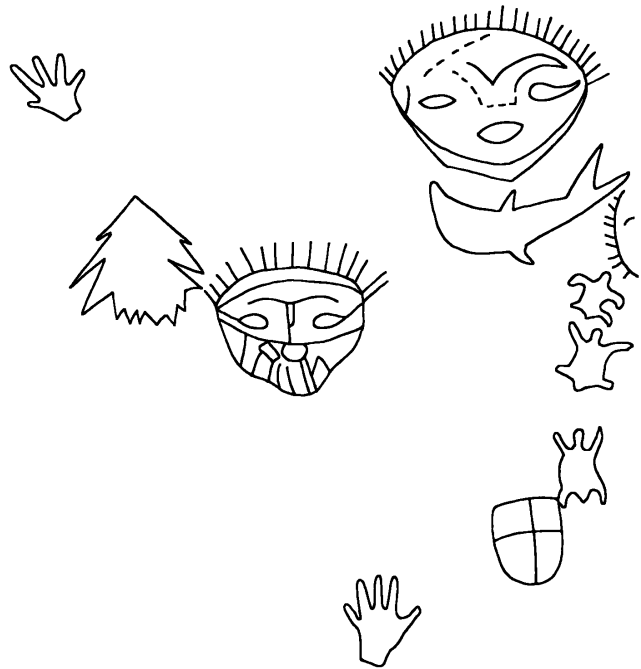

Figure 1.

---

[2] An earlier version of this paper, "The Cult of the Sun in Polynesia", was read to a session of the 20th annual conference of the New Zealand Studies Association together with the Norwegian Maritime Museum and the Kon-Tiki Museum, Oslo in association with the University of South Australia, in Oslo, Norway, on June 26, 2014.

*Tagaloa was the principal god in the archaic West Polynesian beliefs. There were two connected aspects of abundance and fertility in the Polynesian mythology. *Tagaloa was the deity of the sea and the sea creatures. And he was the god of the sun (Steubel 1896). As I have shown, the god *Tini-lau was closely associated with *Tagaloa [*Taga loa] (Rjabchikov 2014). He also had some solar features. The god *Maui-tikitiki was associated with the sun, high sky and fire. Maui was called *Ti'iti'i* in the Samoan mythology (Steubel 1896) and *Titiki Talanga* in the Niuean mythology (Loeb 1926). The name *Maui* reads *Mau i* (Lifted to). The name *Tiki* reads *Tii ki* (Shining to). A Proto-Polynesian rock picture (Vatulele, Fiji) dedicated to the god *Tini-lau is presented in figure 1. The solar signs are the oval (*la'aa*) with a cross as well as three faces (*mata*).

### Data Preserved on Fiji

Let us examine several Fijian stories (Quain 1942; St.-Johnston 1918; Waterhouse 1866; Brewster 1922; Hocart 1929). According to a myth, Tama-ni-geo-loa was the shark deity, 'Tama (Father) of the great shark' [this name could come from the PPN expression *Tama mago loa*]. According to another myth, the spirit Siki-lau married a beautiful woman-spirit. Their two sons once caught a fish. The ghost killed them. According to another myth, the chief Koro-ika was related to fish. According to another myth, *Rogo-waqa* (*Rogo* of the canoe) was the name of a sacred canoe. It played the role of the sea deity.

Tama and Siki-lau are equal to the all-Polynesian gods Tagaloa and Tini-lau or Sini-lau. The nice goddess is Sina. According to another myth, Sina-te-lagi (Sina of the sky) once flew on a great bird. *Koro-ika* (The father of fish; the owner of fish) is the name of the sea god. Rogo is the all-Polynesian divine messenger called *Logo* or *Rogo*.

### Data Preserved on the New Hebrides (Vanuatu)

Let us examine main ideas of the local mythology (Cotterell 1986). The god-creator Tagaro was the husband of Vinmara. His brother was Suqe-Matua. I suggest that these names are PPN reflexes. The name *Tagaro* < *Tagaroa* (*Taga roa*). The goddess Vinmara (< *Fina* [*Fafine*] *Mara*) corresponds to the Rapanui moon goddess Hina Hau Mara. The character Suqe-Matua (Suqe-Father) is equal to the god *Sini-lau. So, the latter form might be initial. Perhaps, the components *Sini (*Tini) and *Lau described the same personage. Also, cf. the Ainu name *Kasunre* < *Ko Sini-rau* (Rjabchikov 2014).

### Data Preserved on Tonga

In compliance with a Tongan myth (Gifford 1924: 16-17), the paramount god Tama-po'uli-ala-ma-foa once ordered the three gods Tangaroa having different epithets to send Lau-fakanaa to the earth. The first name means '*Tama* in the darkness (is) on the path to break the stone.' In a Tahitian creation myth, Ta'aroa (the same Tagaloa) was in a shell initially, and he once left it (Buck 1938). It is clear that Tama-po'uli-ala-ma-foa is equal to this Ta'aroa. As I have predicted in my paper about Tini-rau, *Tama* was the original name of Tangaroa (Rjabchikov 2014; 2012). PAN *t-ama means 'father' (ACD), and PPN *tama means 'ditto' in confidence with Polinsky (Polinskaya 1995: 105, table 18); besides, PPN *tama means 'child' (POLLEX), or 'son' (my reconstruction). At first *Tama was a young and handsome chief-progenitor of a tribe conquered mysterious islands. This Tama or Tangaroa could exist as the trinity, in other words, as the three gods Tangaroa acted simultaneously. The name *Lau-fakanaa* signifying 'Many [objects, perhaps fish] that are hidden' is an early version of the name of the god Tini-lau.

### Tagaloa on the Marquesas Islands

Kirch (1973) has discovered that the shift existed from the marine (wild) to domestic food resources on the base of the Hane Dune site on the Marquesas Islands.

We must agree that the distribution of the pelagic fish during different phases is a real indicator of the possible changes. Table 1 (cf. Dye 1990) contains the valuable pelagic fish taxa at Hane:

Table 1. Number of Identified Specimens (NISP) of the Hane Dune Pelagic Fish

| TAXA | Phases I, II NISP | Phases III, IV NISP |
| --- | --- | --- |
| Scombridae | 14 | 2 |
| Carangidae | 55 | 1 |
| Elasmobranchii | 82 | 14 |
| Belonidae | 4 | 0 |

    A common tendency of the reduction of the number of valuable fish species is appreciable. The role of *Tangaroa* (*Tana'oa* here) as the deity of the ocean and fishery as well as the supreme sun deity decreased gradually. Such a weak character could not be the paramount god. The situation changed. The natives invented the other supreme god, and they called him *Tane* (Male). *Tangaroa* (*Tana'oa*) received the sea and fish only. Maui-tikitiki became Tiki, the first man.

## Marquesan Groups Arrived on Mangareva

The god *Tangaroa* was the father of the gods *Tu*, *Rongo* and some others. *Tiki* was grandson of *Tangaroa*. The god *Tane* was a fisherman. Those Maquesans who were fishermen and preserved their faith in *Tangaroa* as the supreme deity had such beliefs.
    On the other hand, the god *Tiki* met *Hina-One*. The component *One* (Soil, Ground) in this and other cultures denotes the agricultural trend of the development of the civilisation.

## The Sun Deities on New Zealand and Tahiti

The image of Maui-tikitiki was split into two deities: the sun god Tiki was connected with Tane in the official beliefs, and Maui-tikitiki, a folk demigod was outside these ideas. According to Tahitian beliefs, the god Ta'aroa (Tangaroa) was related to the light (Tregear 1891: 29).

## The Colonisation of Easter Island

I suppose that there were several Polynesian waves of immigrants to Rapa Nui. The people of the tribe Tupa-hotu were the descendants of early Marquesan settlers. Tiki or Makemake was their principal deity. It was the sun god.

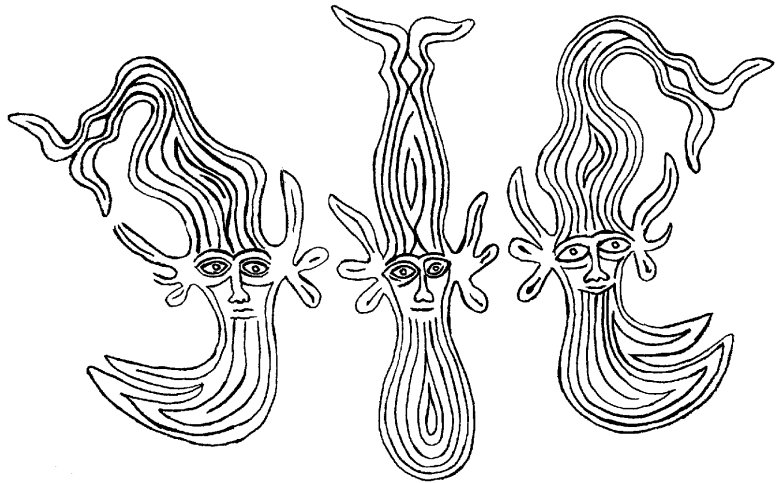

Figure 2.

The people of King Hotu Matua arrived from Mangareva. The god Tangaroa was an ancestor of local kings of the group Miru. This deity had some features of the sun according to the folklore text "Apai" (Rjabchikov 1996: 6).

The three images of the god Tangaroa are depicted on a Rapanui figurine *moai tangata* that is housed in the Peter the Great Museum of Anthropology and Ethnography (Kunstkammer) in St. Petersburg, see figure 2. The similar picture is engraved on a figurine *moai tangata* that is housed in the Néprazjí Museum in Budapest. Another similar picture is engraved on a figurine *moai kavakava* that is housed in the National Museum of Scotland in Edinburgh.

The *rongorongo* record presented on the Mamari tablet (C) is devoted to the god Tamaroa or Tangaroa as the trinity, see figure 3.

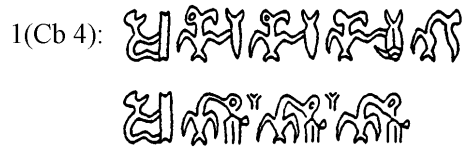

Figure 3.

1 (Cb 4): **6-4 44-16 44-16 44-16 44b 6-4 44-26-15 44-26-15 44-26** *A atua Ta(h)a Kahi, ta(h)a Kahi, ta(h)a Kahi tua, a atua Tamaroa, Tamaroa, Tama.* '(It is) the god 'The shore of the Tuna, the shore of the Tuna, the shore of the Tuna of the open sea', (it is) the god '*Tamaroa = Tangaroa, Tamaroa = Tangaroa, Tama(roa) = Tanga(roa)*'.'

The Juan Haoa and Esteban Atan manuscripts contain the vital information for decoding the *rongorongo* script. According to a legend (Heyerdahl and Ferdon 1965: figures 143-146; the interpretation in Rjabchikov 2010: 39ff), the ceremonial platform Ahu A Tanga played the role of the god Tangaroa. Hence, this Tanga is equal to Tangaroa. Here *a* is an ancient article of personal names. According to another legend (Heyerdahl and Ferdon 1965: figure 129; cf. Barthel 1978: 4), Nga Tavake, Te Ohiro and Hau, the descendants of Te Taanga, arrived together or separately on Easter Island. Since the god Tangaroa was an ancestor of the local kings (Métraux 1940: 127), this Te Taanga is equal to Tangaroa. Here *te* is an ancient article of personal names, too. On the basis of these data, one can read two fragments of the texts on the Santiago staff (I) that belonged to King Nga Ara, see figure 4.

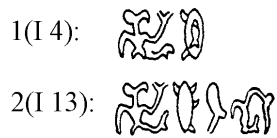

Figure 4.

1 (I 4): **44-(102)-28 12** *Taanga ika* 'Tangaroa, (the god) of the fish.'
2 (I 13): **44-(102)-28 44b (102)** *Taanga tua* 'Tangaroa, (the god) of the open sea.'

**The Report about a Solar Eclipse on the Santiago Staff**

Consider the following record put down on the staff, see figure 5. Here and everywhere else, I use the computer program RedShift Multimedia Astronomy (Maris Multimedia, San Rafael, USA) to look at the heavens above Easter Island.

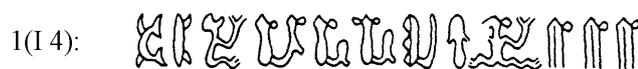

Figure 5.

1 (I 4): **8 28 6 56 (102) 56 (102) 56 (102) 18-4 22 69 26-26-26** *Matua nga TANGATA* [= *Koro*] *po-po-po te atua rapa Moko maa-maa-maa.* '(It was) the month *Koro* [Father literally; December chiefly]. (It was) the long night because of (the appearance of) the paddle of the Lizard (*Hiro*) during the very bright sun.'

    Cf. Rapanui *matu'a* 'father' and Old Rapanui *koro* 'ditto.' Old Rapanui *nga* means 'progenitor,' cf. Mangarevan *nana* 'to create; to produce; to grow' (Rjabchikov 2014).

    It is the description of the partial solar eclipse of December 20, 1805 A.D. during the sunset. Thus, King Kai Makoi the First died before that day (the text about that event is written on the first line of the staff). In 1805 A.D. King Nga Ara was around 25 years old, and he began to rule.

### Several Words about the Mataveri Calendar

On a rock that has been found at Mataveri (an important area of the bird-man cult) some lines were incised; most of them were the directions of the setting sun in compliance with Liller (1989). I have determined the corresponding days for the year 1775 A.D., see table 2.

Table 2. The Dates Calculated

**June 22** (the azimuth of the sun = 296.2°): one day after the winter solstice;
July 21 (292.5°): the star Capella (α Aurigae) before dawn;
August 11 (286.7°): the star Pollux (β Geminorum) before dawn;
September 2 or 3 (277.9°): the star β Centauri [*Nga Vaka*] before dawn;
**September 21** (270.1°): the day before the vernal equinox, the key moment of the bird-man feast;
September 24 (268.7°): the new moon;
September 27 (267.4°);
October 1 (265.9°);
October 3 (264.7°);
October 22 (256.8°): near the new moon;
November 8 (250.7°): the star Spica (α Virginis) before dawn;
November 12 (249.3°);
November 14 (248.7°);
November 23 (246.3°): the new moon;
**December 20** (the azimuth of Aldebaran = 339.1°): the star Aldebaran (α Tauri) at night;
**December 21** (the azimuth of Aldebaran = 322.1°; the azimuth of Canopus = 177.5°): the stars Aldebaran (α Tauri) and Canopus (α Carinae) on the same night (Rjabchikov 2013: 7); the day of the summer solstice.

    One can decipher the parallel text on the Aruku-Kurenga tablet (B), see figure 6.

1(Bv 4): 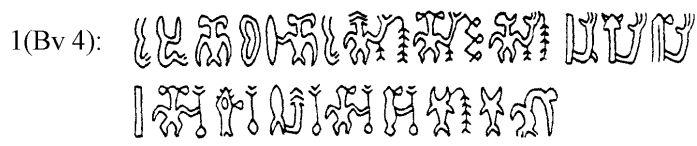

Figure 6.

1 (Bv 4): **43 2 44 47 30-44 43 33 6-15 24 6-15 52 6-15 24 4-15 21-15 26-15 4-6 101 56 101 12 4-33 101 6 101 56 101 11 24 11 19** *Ma Hina Ta(h)a avae Anakena, ma ua, Hora ari, Hora iti, Hora ari. Atua roa Koro Maro tuha. O(h)o po, o(h)o Ika Atua, o(h)o Ha, o(h)o po, o(h)o Mango ari, Mango ki.* 'The moon (*Rongo*) *Tane* of the month *Anakena* [July chiefly] goes, the rains, (then the month) *Hora* [*Hora-Nui*; September chiefly] of the bright sun after (the month) *Hora-Iti* [August chiefly], (i.e. the month) *Hora* [*Hora-Nui*; September chiefly] of the bright sun go. The great god (the month) *Koro* [December chiefly] (came) (after) the time interval (month) *Maro* [June chiefly]. The nights enter, the moon *Atua* enters, the constellation Auriga enters, the nights enter, the star Pollux (β Geminorum) (?) enters.'

## On Some Local Place Names Related to the Sun

Two mountains situated on the Poike peninsula on the eastern part of Rapa Nui, Puakatiki and Maunga Teatea, are indicators of the rising sun (cf. Liller 1991). The first place name reads *Pua Ka Tiki* 'The summit of the fire of Tiki.' The second place name means 'The mountain of the white (= bright) colour.' According to a Rapanui legend (Felbermayer 1963), a priest once stood on the Tongariki platform and prayed. He wanted the rain to fall down. He shouted: '*Tiki*, hide the face!' Hence, Tiki was the sun deity in fact. Here Rapanui *api* (to hide) was used. So, the place names *Apina-Iti* (The Small Apina) and *Apina-Nui* (The Big Apina) corresponding to the western shore of the island denote the western direction. The term *Apina* derived from Rapanui *Apinga* (Hidden; perhaps the setting sun).

In the eastern part of the island not far from the platform Ahu Hanga Hoonu one can find the following platforms: Ahu Tare named after son of the god Tiki, Ahu Hanga o Miti (cf. Rapanui *miti* 'to dry up'), and Ahu Oho Tea (cf. Rapanui *oho tea* [the whiteness enters literally], *otea* 'dawn,' *ootea* [*oho tea*] 'to dawn'). These spots could be relevant to archaic sun observatories.

## Conclusions

On the basis of the Tongan and Tahitian folklore the contents of the myth about the creation of the Universe have been reconstructed. The early name of Tagaloa (Tangaloa, Tangaroa etc.) was *Tama* (Chief, Man, Father). He was the principal deity in Western Polynesia. This character was the god of fishermen, too. Also, he was the god of the sun. Tagaloa (Tanga loa) or Tini-lau was incarnated into big sea creatures (shark, whale, seal etc.). Maui was called *Ti'iti'i* in the Samoan mythology and *Titiki Talanga* in the Niuean mythology. He got the fire and uplifted the heavens above the ground highly. One can insist that in the distant past Tama was a young tall chief, the head of the group of explorers. The name *Maui* (*\*Mau i*) denotes the elevation of the sky; and the reduplicated name *Tiki* (*\*Tii ki*) denotes the solar rays. The presence of the human beings on the Marquesas Islands (Eastern Polynesia) broke the religious concepts of the early voyagers. The excavations demonstrate that the number of the caught valuable pelagic fish decreased dramatically. In that situation Tangaroa could not stay in the role of the paramount deity. He was turned into the god of the fishermen only. The natives invented another anthropomorphic god named *Tane* (Man). The image of the archaic Western Maui-tikitiki was split into two gods. First, Tiki, the first man, was invented. Second, the same Maui-tikitiki became the demigod of commons, he was outside the religious cult. These data allow deciphering differences between religious beliefs of some peoples in Eastern Polynesia. Several Rapanui place names have been decoded. This paper is suggested as a contribution for understanding the Polynesian ethno-archaeology and linguistics. Some data on the Easter Island archaeoastronomy are offered, too.

# APPENDIX 1

The articles of mine, Rjabchikov 2014 and Rjabchikov 2012, would be read in this sequence. The first manuscript was ready before the preparation of the second manuscript.

# APPENDIX 2

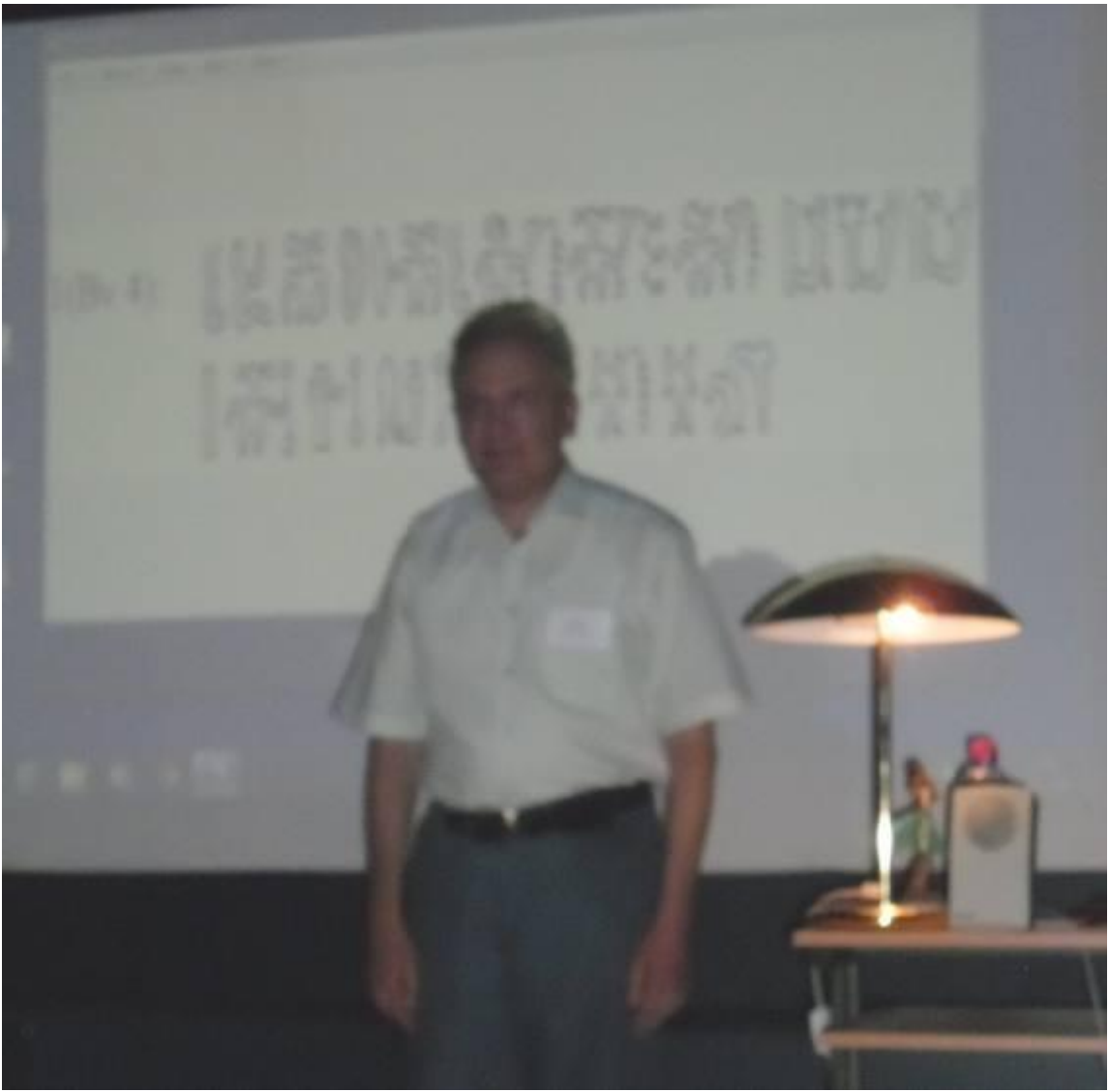

I had delivered my paper.
I was warmly accepted by the
participants of the conference.